\documentclass[a4paper,11pt]{article}
\pdfoutput=1

\usepackage{jheppub} 

\usepackage[T1]{fontenc}

\usepackage{amsmath}
\usepackage{multirow}
\usepackage{booktabs}
\usepackage{multirow}

\usepackage{graphicx}

\usepackage{caption}

\usepackage{subcaption}

\usepackage{float}

\newcommand{\be}{\begin{eqnarray}}
\newcommand{\ee}{\end{eqnarray}}

\def\p{\pi}

\def\nn{\nonumber}
\def\p{\partial}
\def\ls{\left[}
\def\rs{\right]}
\def\lc{\left\{}
\def\rc{\right\}}

\newcommand{\bi}{\begin{itemize}}
\newcommand{\ei}{\end{itemize}}

\title{Naturalness and Chaotic 
Inflation in Supergravity   \\ 
from 
Massive Vector Multiplets}

 \author{Fotis Farakos}
 \author{and Rikard von Unge}
 \affiliation{ Institute for Theoretical Physics, Masaryk University, \\
611 37 Brno, Czech Republic}

\emailAdd{fotisf@mail.muni.cz}
\emailAdd{unge@physics.muni.cz}

\abstract{We study the embedding of the quadratic model of chaotic 
inflation into the 4D, N=1 minimal theories of supergravity by the use of massive vector multiplets and 
investigate its robustness against higher order corrections. 
In particular, we investigate the criterion of  technical naturalness for the inflaton potential.   
In the framework of the new-minimal formulation 
the massive vector multiplet is built in terms of a real linear multiplet coupled to a vector multiplet 
via the 4D analog of the Green-Schwarz term. 
This theory gives rise to a single-field quadratic model of chaotic inflation, 
which is protected by an shift symmetry which naturally suppresses the higher order corrections. 
The embedding in the old-minimal formulation is again achieved 
in terms of a massive  vector multiplet and also gives rise to single-field inflation.  
Nevertheless in this case there is no obvious symmetry to protect the model from higher order corrections. 
}

\arxivnumber{1404.3739}

\begin{document} 
\maketitle
\flushbottom

\section{Introduction}

There has been stimulating progress in observational inflationary cosmology. 
The  BICEP2 experiment \cite{Ade:2014xna} has announced the observation of the ratio of tensor to scalar perturbations
of the metric to be
\be
r  = 0.2^{+0.07}_{-0.05} 
\ee
which has triggered a discussion \cite{Kehagias:2014wza,Ibanez:2014zsa,Dvali:2014ssa,Lyth:2014yya,
Kaloper:2008fb,Kaloper:2014zba} of the implications of these findings on 
theoretical models for physics beyond the electroweak scale.

The simplest model which is favored by the combined PLANCK and BICEP2 \cite{Ade:2014xna,Ade:2013uln} data  is the quadratic model 
of chaotic inflation \cite{Linde:1983gd}, with potential 
\be
\label{chaotic}
{\cal V} = \frac{1}{2} m^2 \phi^2  
\ee
which for $50$ 
e-foldings of inflation gives rise to a ratio of tensor to scalar perturbation of $r \approx 0.16$.

The new experimental results and the fact that inflation is a high energy process gives us a new
possibility to study models of physics beyond the Standard Model. In particular, any serious candidate
for new physics will have to be able to reproduce these results. 
In this work we will focus on supersymmetric theories. 
It is then important to embed the chaotic model of inflation (\ref{chaotic}) into 
the theory of supergravity \cite{Lyth:1998xn,Linde:2007fr}.

An embedding into old-minimal supergravity was carried out successfully in Ref. \cite{Kawasaki:2000yn}, 
by employing chiral multiplets and was further discussed in Ref. \cite{Demozzi:2010aj}.   
A general chaotic inflation was also discussed in Ref. \cite{Kallosh:2010ug,Kallosh:2010xz}, 
where these proposals are easier to embed into phenomenological models. 
An interesting attempt to give a geometric origin to a chaotic inflationary phase in supergravity   
was initiated recently in Ref. \cite{Ferrara:2014ima}, 
again involving two chiral multiplets in the dual picture, 
and further discussed in Ref. \cite{KLr,El,jap}.

Another interesting possibility is to study models with massive vector multiplets, 
first coupled to the old-minimal supergravity in Ref. \cite{Mukhi:1979wc,VanProeyen:1979ks}, 
where also the quadratic potential was initially discussed \cite{VanProeyen:1979ks}. 
The relation of massive vector multiplets to new-minimal higher derivative supergravity was 
first pointed out in Ref. \cite{Cecotti:1987qe}. 
In a series of papers \cite{Farakos:2013cqa,Ferrara:2013rsa,Ferrara:2013kca} 
single-field inflationary models utilizing 
massive vector multiplets and possible higher derivative corrections were systematically studied, 
and a single-field chaotic model was introduced in Ref. \cite{Ferrara:2013rsa}. 
Moreover, the gauged isometries of  the minimal supergravity models of inflation 
were investigated  in Ref. \cite{Ferrara:2013eqa,Ferrara:2014rya}.  
Quadratic chaotic inflationary models where the D-term dominates over the F-term 
were studied in Ref. \cite{Kadota:2007nc,Kadota:2008pm}.  
Finally, a different perspective on chaotic inflation from D-terms 
in supergravity may be found in Ref. \cite{Dalianis:2014sqa}.

In general, 
the potential (\ref{chaotic}) is not straightforward to reproduce.
The usual issues one encounters are 
\begin{itemize}

\item Identify the one and only scalar which drives inflation. 

\item Stabilize the other scalars, and explain why they do not ruin inflation. 

\item Higher order corrections may spoil inflation; the notorious $\eta$-problem. 

\end{itemize}
In this work we investigate the possibility of  embedding the quadratic model of inflation in supergravity, 
with the use of massive vector multiplets and we will see how the aforementioned issues are addressed. 
In particular, a generic property of inflationary models utilizing massive vector multiplets 
is that there is no need to stabilize any additional fields nor identify the inflaton; 
these are by construction single-field models, and thus the first two aforementioned issues are automatically solved. 
To address the issue of higher order corrections we will rely on the existence of a softly broken shift symmetry 
and invoke technical naturalness in the sense of  't Hooft \cite{'tHooft:1979bh}.

\section{Chaotic inflation in new-minimal supergravity}

Let us start with a real linear multiplet, 
and couple it to the new-minimal supergravity \cite{Sohnius:1981tp,Sohnius:1982fw,Ferrara:1988qxa,Ovrut:1988fd}. 
The definition of the real linear superfield in this framework is
\be
\label{new}
\nabla^2 L = \bar \nabla^2 L = 0 . 
\ee
The definitions of the bosonic components are   
\be
L | = \phi \ \  , \ \ -\frac{1}{2} [\nabla_\alpha , \bar \nabla_{\dot \alpha} ] L | = h_{\alpha \dot \alpha} 
\ee
with 
\be
h_m = - \frac{1}{2} \epsilon_{mnrs} \p^n b^{rs} - 2 \phi H_m
\ee
where $b_{mn}$ is the two form of the real linear multiplet and $H_m$ 
is an auxiliary field of the new-minimal supergravity formulation which we will review later. 
It is easy to verify that the minimal kinematic term 
\be
\label{kin}
- \int d^4 \theta E L^2 = - \frac{1}{2} e \p \phi \p \phi + \frac{1}{2} e h_m h^m 
\ee
allows a  shift symmetry for the superfield 
\be
\label{shift}
L \rightarrow L + c M_P  
\ee
for some real constant $c$, which translates into
 \be
\label{shiftphi}
\phi \rightarrow \phi + c M_P  
\ee
for the real scalar lowest component. 
We would like to stress that the shift (\ref{shift}) is also a symmetry of (\ref{new}) and thus 
does not violate the definition of the real linear multiplet.

The shift symmetry (\ref{shift}) protects the minimal kinematic term from higher order corrections 
and will later shield the model against the $\eta$-problem. 
For example the possible higher order correction
\be
\nn
\label{L4}
-\frac{1}{(\text{some scale})^4}\int d^4 \theta E L^4 \rightarrow \text{violates symmetry}  
\ee
is ruled out since the quartic term ($L^4$) violates the shift symmetry.

Of course, this exact shift symmetry (\ref{shift})  rules 
out the possibility of introducing a potential for the theory. 
Here is where the Green-Schwarz term comes in. 
It is well established that a gauge anomaly can be canceled by introducing a two-form which
couples to the gauge field and gives rise to tree diagrams that cancel the anomalous loop diagrams. 
This mechanism exists also in supergravity theories and in four dimensions the coupling of the two-form
with the gauge field is given by the gauge invariant contact term
\be
\begin{split}
\label{GS}
- g M \int d^4 \theta \,E \, L \, V  = - \frac{1}{2} e g M \phi \text{D}  
+ \frac{1}{2} e g M v_m \left( h^m +  2 \phi  H^m \right) 
\end{split}
\ee
where $V$ the vector superfield of the would-be-anomalous $U(1)$, 
and $L$ a real linear multiplet containing the two-form. 
The fields $\text{D}$ and $v_m$ are the auxiliary real scalar 
and the physical vector  of the $U(1)$ vector superfield.
In  Ref. \cite{CFG87} it was shown that the effect of this term is to introduce a realization of the 
Stueckelberg mechanism where the two-form of the real linear is eaten by the vector. 
The Green-Schwarz mechanism was further investigated in minimal supergravity in Ref. \cite{Lopes Cardoso:1991zt}. 
It is in fact a way to write down massive vector multiplets \cite{Siegel:1979ai,CFG87}. 
On the other hand this same term (\ref{GS}) violates the shift symmetry (\ref{shift}) and as we will see it creates a potential. 
Thus the small breaking of the shift symmetry 
is generated by the Green-Schwarz term 
and it is expected that as long as 
\be
\label{gmh}
g M << H 
\ee  
where $H$ is the Hubble constant, 
the effect of the small breaking on the kinematic term (\ref{kin}) is negligible. 
This is  natural  in the 't Hooft sense since for  $g M \rightarrow 0$ the shift symmetry is restored, 
and the symmetry of the system is enhanced \cite{'tHooft:1979bh}. 
To better understand this argument it is convenient to think of the mass $M$ as 
parameterizing the flow of the theory through some parameter space. 
When the theory sits on the $M_*=0$ point, all the operators that violate the symmetry vanish. 
Far away from the special point $M_*=0$, the symmetry violating operators become large. 
Thus, for a small value of $M$ close to $M_*=0$ the operators that violate the shift symmetry have to be highly suppressed. 
More specifically for example
\be
\nn
-\frac{1}{(\text{some scale})^4}\int d^4 \theta E L^4 \rightarrow \text{naturally suppressed} . 
\ee
For similar considerations in the old-minimal supergravity framework, 
in a model of two chiral multiplets,  see Ref. \cite{Kawasaki:2000yn}.

Taking this into account we may proceed to investigate the model in more detail. 
As we have mentioned, our interest now lies in the new-minimal supergravity \cite{Sohnius:1981tp}, 
which originates from the superconformal supergravity after appropriate 
gauge fixing \cite{Kugo:1982cu,Ferrara:1983dh,Gates:1983nr,Buchbinder:1995uq,Butter:2009cp}. 
Note that the new-minimal supergravity allowes only R-invariant Lagrangians. 
The bosonic sector of the pure theory reads
\be
-2  \int d^4 \theta \,E \, V_{\text{R}}  = \frac{1}{2}  e \left( R + 6 H_m H^m \right)  + 2  e A^-_m H^m . 
\ee
On top of the graviton $e^a_m$ and the gravitino $\psi_m^\alpha$, 
this theory contains two auxiliary fields: 
the $A_m$ which gauges the  R-symmetry and the two-form $B_{mn}$, 
which only appears through the dual of its field strength $H^m$. 
Let us mention that 
\be
A^-_m = A_m -3 H_m.
\ee 
The properties of this minimal supergravity were investigated in a series of papers  \cite{Sohnius:1981tp,Sohnius:1982fw,Ferrara:1983dh,CFG87,Cecotti:1987qe,Cecotti:1987mr,
Ferrara:1988pd,Ferrara:1988qxa,Ovrut:1988fd,Ovrut:1989bh,Ovrut:1990nk}.

For the total Lagrangian to be manifestly supersymmetric we will write it down 
in new-minimal supergravity superspace \cite{Ferrara:1988qxa}, 
and then turn to component form. 
The theory we wish to consider is 
\be
\begin{split}
\label{A1}
{\cal L}=
& 
-2 M_P^2 \int d^4 \theta \,E \,  V_{\text{R}} 
+ \frac{1}{4} \ls \int d^2 \theta  \,  {\cal E} \, W^2(V) +h.c. \rs
\\
&
-g M  \int d^4 \theta  \,E \, L V
- \int d^4 \theta  \,E \, L^2
\end{split}
\ee
which as we mentioned has a matter sector combination of real linear superfield $L$ 
and vector superfield $V$ that reproduce the Green-Schwarz mechanism  in supergravity \cite{CFG87}. 
Here
\be
 W_{\alpha}(V) = - \frac{1}{4} \bar \nabla^2 \nabla_{\alpha} V
\ee
is the standard field strength chiral superfield. 
For the pure gauge sector we have the bosonic components  
\be
\begin{split}
 \frac{1}{4 } \int d^2 \theta  \,  {\cal E} \, W^2(V) + c.c. = 
 - \frac{1}{4 } e F^{mn} F_{mn} (v) + \frac{1}{2 } e \text{D}^2 
\end{split}
\ee
for 
\be
F_{mn}(v) = \p_m v_n - \p_n v_m .
\ee

Now we can find the full bosonic sector of (\ref{A1}) 
\be
\label{A2}
\begin{split}
e^{-1} {\cal L}^B  = &  \frac{1}{2} M_P^2 \left( R + 6 H_m H^m \right)  + 2 M_P^2 A^-_m H^m  
\\
& - \frac{1}{2}  \p \phi \p \phi + \frac{1}{2}  h_m h^m 
 + \frac{1}{2 }  \text{D}^2 - \frac{1}{4 }  F^{mn} F_{mn} 
\\
& + \frac{1}{2} g M  v_m \left( h^m +  2 \phi  H^m \right) 
- \frac{1}{2} g M \phi \text{D}  .
\end{split}
\ee
Let us integrate out the supergravity auxiliary and the $h$-field. 
First we make $h$ and $H$  unconstrained by introducing Lagrange multipliers $X$ and $Y$
\be
\begin{split}
e^{-1} {\cal L}_{aux} =&   3 M_P^2 H_m H^m  + 2 M_P^2 A^-_m H^m  + \p_n X  H^n 
\\
& - \frac{1}{2} g M  \phi \text{D}  + \frac{1}{2} g  M v_m \left( h^m +  2 \phi  H^m \right) 
\\
& + \frac{1}{2}  \text{D}^2 
+ \frac{1}{2}  h_m h^m  + \p_m Y (  h^m +  2 \phi  H^m ) .
\end{split}
\ee
{}From the equations of motion of $A^-_m$ we find the condition
\be
\label{Heq}
  H^m = 0
\ee
and from the $H_m$ equations we find
\be
\begin{split}
6 M_P^2 H_m  + 2 M_P^2 A^-_m 
 +gM v_m  \phi  
+ \p_m X   + 2 \phi  \p_m Y     = 0
\end{split}
\ee
which combined with (\ref{Heq}) leads to
\be
\label{A-}
A^-_m   = 
 - \frac{gM v_m  \phi  
+ \p_m X   + 2 \phi  \p_m Y  }{( 2 M_P^2) } .
\ee
In fact the equation (\ref{A-}) never shows up, except in the supersymmetry transformations 
of the on-shell theory. 
The auxiliary Lagrangian becomes 
\be
\begin{split}
e^{-1} {\cal L}_{aux} =
 - \frac{1}{2} g M \phi \text{D}  + \frac{1}{2}g M v_m  h^m 
 + \frac{1}{2}  \text{D}^2 
+ \frac{1}{2}  h_m h^m   + \p_m Y  h^m  . 
\end{split}
\ee
Integrating out $h^m$ and D we find
\be
e^{-1} {\cal L}_{aux} =  
- \frac{1}{8} (g M v_m +2 \p_m Y )^2
 -  \frac{g^2}{ 8} M^2 \phi^2.
\ee

The model (\ref{A1}) after integrating out all the non-propagating fields  is 
\be
\label{fin}
\begin{split}
e^{-1} {\cal L}^B =  
\frac{1}{2}  M_P^2  R  - \frac{1}{2}  \p \phi \p \phi -  \frac{g^2}{ 8} M^2  \phi^2 
- \frac{1}{4}  F^{mn} F_{mn} (v) 
- \frac{1}{8} g^2 M^2 v^m v_m  
\end{split}
\ee
where we have shifted
\be
\label{stueckelberg}
v_m \rightarrow   v_m + \frac{2}{g M} \p_m Y . 
\ee
It is clear from (\ref{stueckelberg}) that the Stueckelberg mechanism is at work.

The sector relevant to inflation  reads
\be
e^{-1} {\cal L}_{scalar} =  \frac{1}{2}  M_P^2  R   - \frac{1}{2}  \p \phi \p \phi 
-  \frac{m^2}{ 2}   \phi^2 
\ee 
where we have replaced
\be
g M = 2 m  
\ee
which is fixed by the  observational data (see for example \cite{Lyth:1998xn}) to be  
\be
m \sim 10^{13} GeV  . 
\ee
During inflation since the $\eta$ slow-roll parameter is small  we see that indeed 
\be 
\frac{g M}{H} \sim \frac{M_P}{\phi} << 1 
\ee  
and relation (\ref{gmh}) holds.

We see that the model (\ref{A1}) successfully reproduces the simplest model of chaotic inflation (\ref{chaotic}). 
Moreover there is no ambiguity in choosing the inflaton field, 
which is a common issue in supergravity inflation. 
The ambiguity is resolved by the simple fact that there is no other scalar in the first place. 
Indeed, thanks to the Stueckelberg mechanism the second scalar of the inflaton multiplet is eaten by the vector field. 
Thus we have a   model of single-field chaotic inflation  in supergravity  which is technically natural
 employing a softly broken  shift symmetry.

We should mention that single-field inflationary models with the use of massive vector multiplets 
have been introduced only recently in the literature \cite{Farakos:2013cqa,Ferrara:2013rsa,Ferrara:2013kca}, 
and in particular in Ref. \cite{Ferrara:2013rsa} a very interesting discussion on their properties can be found, 
including a realization of the chaotic model. 
In fact the model studied here is dual to the chaotic models of  Ref. \cite{Ferrara:2013rsa}, 
and thus essentially reproduces equivalent results. 
On the other hand the "linear superfield - vector superfield" picture we presented makes 
the  technical naturalness of the 
model manifest in the new-minimal supergravity formulation.

As we have mentioned earlier, 
the Lagrangian (\ref{A1}) describes a massive vector multiplet. 
This may be seen by rewriting (\ref{A1}) as 
\be
\begin{split}
\label{A2}
{\cal L}=& -2 M_P^2 \int d^4 \theta \,E \,  V_{\text{R}} 
-g M  \int d^4 \theta  \,E \, L (V + \frac{1}{g M} \Phi +  \frac{1}{g M} \bar \Phi)
\\
&
+ \frac{1}{4} \ls \int d^2 \theta  \,  {\cal E} \, W^2(V) +h.c. \rs
- \int d^4 \theta  \,E \, L^2
\end{split}
\ee
where now $L$ is unconstrained. 
{}From (\ref{A2}) we see that the chiral superfield $\Phi$ has to carry a vanishing R-charge 
for the chiral-linear duality to be possible. 
Then by integrating out $L$ from (\ref{A2}) we have 
\be
\begin{split}
\label{A3}
{\cal L}=& -2 M_P^2 \int d^4 \theta \,E \,  V_{\text{R}} 
+ \frac{1}{4} \ls \int d^2 \theta  \,  {\cal E} \, W^2(V) +h.c. \rs
+ \frac{1}{4} g^2 M^2  \int d^4 \theta  \,E \, V^2  
\end{split}
\ee
where we have shifted 
\be
V \rightarrow V - \frac{1}{g M} \Phi -  \frac{1}{g M} \bar \Phi .
\ee 
The Lagrangian (\ref{A3}) describes a massive vector multiplet coupled to the new-minimal supergravity. 
It is easy to see that the bosonic sector of (\ref{A3}) will be given again by (\ref{fin}). 
In the limit $gM \rightarrow 0$ gauge invariance is restored, 
and the theory will be described by a massless vector multiplet 
and a massless chiral multiplet. 
Indeed in this limit the theory will become 
\be
\begin{split}
\label{A33}
{\cal L}=& -2 M_P^2 \int d^4 \theta \,E \,  V_{\text{R}} 
+ \frac{1}{4} \ls \int d^2 \theta  \,  {\cal E} \, W^2(V) +h.c. \rs
+ \frac{1}{2}  \int d^4 \theta  \,E \, \bar \Phi \Phi   
\end{split}
\ee
and the shift symmetry will translate into
\be
\label{dd}
\Phi \rightarrow \Phi + d \, M_P . 
\ee
In (\ref{dd}) the constant $d$ can be complex. 
Note that (\ref{A2}) allowed only for a pure imaginary shift of the chiral superfield 
and thus only in the $gM \rightarrow 0$ limit $d$ can be complex. 
The fact that the shift (\ref{dd}) is a symmetry of (\ref{A33}) is connected to the 
structure of the new-minimal supergravity, which gives 
\be
\int d^4 \theta  \,E \, \ \Phi   = \int d^4 \theta  \,E \, \ \bar \Phi = 0 
\ee
for a chiral superfield with vanishing R-charge.

If we instead start with the most general (up to two derivatives), 
gauge invariant coupling of the real linear with the vector multiplet,  
the superspace Lagrangian reads
\be
\begin{split}
\label{A11}
{\cal L}=& -2 M_P^2 \int d^4 \theta \,E \,  V_{\text{R}} 
+ \frac{1}{4} \ls \int d^2 \theta  \,  {\cal E} \, W^2(V) +h.c. \rs
\\
&
-g M  \int d^4 \theta  \,E \, L V
- \int d^4 \theta  \,E \, {\cal F}(L)
\end{split}
\ee
and after integrating out all the auxiliary field sector we find the bosonic part 
\be
\label{fin2}
\begin{split}
e^{-1} {\cal L}^B =   \frac{1}{2}  M_P^2  R  - \frac{1}{4} {\cal F}'' \p \phi \p \phi -  \frac{g^2}{ 8} M^2  \phi^2 
- \frac{1}{4}  F^{mn} F_{mn} (v) 
- \frac{1}{4 {\cal F}''} g^2 M^2 v^m v_m  
\end{split}
\ee
where 
\be
{\cal F}''(\phi) = \frac{\p^2 {\cal F}(\phi)}{\p \phi \p \phi}  . 
\ee 
Again in (\ref{fin2}) the vector has eaten the Lagrange multiplier $Y$.

For a general kinematic function 
\be
\label{FF}
{\cal F}(\phi) = c_0 +  c_2 \phi^2 + c_3 \phi^3 + \dots 
\ee
when all the higher order terms are present  there is no shift symmetry, 
and in general ${\cal F}(\phi)$ will receive large corrections.  
Only in the case when 
\be
c_n = 0  \ \ , \ \ n\ge 3
\ee
does $\cal F$ respect the shift symmetry and any correction to the higher order terms will have to be generated 
by the Green-Schwarz term and thus be  suppressed.

\section{Chaotic inflation in old-minimal supergravity}

Now we investigate the embedding of the 
quadratic chaotic model in the old-minimal superspace \cite{Wess:1992cp}. 
This supergravity also originates from the superconformal supergravity after appropriate 
gauge fixing \cite{Kugo:1982cu,Ferrara:1983dh,Gates:1983nr,Buchbinder:1995uq,Butter:2009cp}. 
A complete treatment of the curvature superfields of this theory can be found in Ref. \cite{Ferrara:1988qx}. 
In addition to the graviton and the gravitino, the pure theory contains two auxiliary fields: 
a complex scalar $u$, and a real vector $b_m$. 
It is well known how to couple a self-interacting massive vector multiplet 
to the old-minimal supergravity \cite{Mukhi:1979wc,VanProeyen:1979ks}.

Following the results of the previous section, 
we consider the superspace Lagrangian 
in the old-minimal formulation  
\be
\begin{split}
\label{B1}
{\cal L}=& - 3 M_P^2 \int d^2 \Theta \,2 {\cal E} \,  {\cal R} + h.c. 
+ \frac{1}{4} \ls \int d^2 \Theta  \,  2{\cal E} \, W^2(V) +h.c. \rs
\\
&
-g M  \int d^4 \theta  \,E \,  {\cal Q} V
- \int d^4 \theta  \,E \, {\cal G}( {\cal Q})  
\end{split}
\ee
where $\cal Q$ is a real linear multiplet with definition \cite{Binetruy:2000zx} 
\be
\label{oldL}
(\bar {\cal D}^2 - 8 {\cal R}) {\cal Q} = 0 . 
\ee
We see from (\ref{oldL}) that the shift symmetry  argument used in the new-minimal case does not apply,  
since the shift 
\be
{\cal Q} \rightarrow {\cal Q} + c M_P
\ee
for a real constant $c$, violates the definition (\ref{oldL}). 
Thus a choice of a quadratic kinematic function
\be
\label{Q^2}
{\cal G}( {\cal Q}) = {\cal Q}^2  
\ee
is not protected by a shift symmetry.

To find the component form and relate to the known results, 
it is better to rewrite the Lagrangian (\ref{B1}) as the coupling of a massive vector 
multiplet to supergravity. 
For the part containing the real linear superfield we have
\be
\begin{split}
\label{B2}
{\cal L}_{\cal Q}  =&  -g M  \int d^4 \theta  \,E \,  {\cal Q} V
- \int d^4 \theta  \,E \, {\cal G}( {\cal Q}) 
\\
 = &  -g M  \int d^4 \theta  \,E \,  {\cal Q} (V + \frac{\Phi}{g M} + \frac{\bar \Phi}{g M} )
- \int d^4 \theta  \,E \, {\cal G}( {\cal Q}) . 
\end{split}
\ee
In (\ref{B2}) the superfield $\cal Q$ is unconstrained, 
and it may be integrated out via the equations of motion
\be
 {\cal G}'( {\cal Q}) =- g M  (V + \frac{\Phi}{g M} + \frac{\bar \Phi}{g M} ) . 
\ee
The theory then becomes
\be
\label{oldminZ}
\begin{split}
{\cal L} =  \frac{1}{4} \int d^2 \Theta \, 2 {\cal E} \, W^2(V) + c.c. 
 +  \int d^2 \Theta \, 2 {\cal E} \lc  -\frac{1}{8} ( \bar {\cal D}^2 -8 {\cal R} ) {\cal Z}(V)  \rc +c.c. 
\end{split}
\ee
with 
\be
{\cal Z}(V) = -3 M_P^2 - \ls  {\cal G}( {\cal Q}) +g M {\cal Q} V \rs_{ {\cal G}'( {\cal Q})=-gMV} . 
\ee
The Lagrangian (\ref{oldminZ}) is  defined for 
any hermitian function $ {\cal Z}(V)$ of dimension $ [{\cal Z}(V)]=2$ of the real vector superfield $V$.

To turn to component form let us first define the components of the massive vector multiplet as
\be
\begin{split}
C &=V| 
\\
N &= -\frac{1}{4} {\cal D}^2 V|  
\\
v_{\alpha \dot \alpha} &= - \frac{1}{2} [{\cal D}_\alpha , \bar {\cal D}_{\dot \alpha} ] V|  
\\
\text{D}  &= \frac{1}{8} {\cal D}^\alpha  ( \bar {\cal D}^2 -8 {\cal R} ) {\cal D}_\alpha \, V| . 
\end{split}
\ee
The bosonic sector of (\ref{oldminZ}) reads 
\be
\label{compZ}
\begin{split}
e^{-1} {\cal L}^B = & - \frac{1}{4} F^{mn} F_{mn} (v) +\frac{1}{2} \text{D}^2 -\frac{1}{4}  {\cal Z}'' b^m b_m
 -\frac{1}{3}  {\cal Z}' \bar u \bar N  -\frac{1}{3}  {\cal Z}' u N  + \frac{1}{9}  {\cal Z} u \bar u 
\\
&+ \frac{1}{6}  {\cal Z} R +  {\cal Z}'' N \bar N -\frac{1}{9}  {\cal Z} b^m b_m
 + \frac{1}{2}  {\cal Z}' \text{D} - \frac{1}{4}  {\cal Z}'' \p C \p C +\frac{1}{3}  {\cal Z}' b^m v_m   
\end{split}
\ee
where now ${\cal Z}$ is a function of the lowest component $C$ and 
\be
{\cal Z}' = \frac{\p {\cal Z}}{\p C} \ \ , \ \ {\cal Z}'' = \frac{\p^2 {\cal Z}}{\p C \p C} . 
\ee
After integrating out the auxiliary sector and performing the appropriate Weyl rescalings one finds \cite{VanProeyen:1979ks} 
\be
\label{OLD}
\begin{split}
e^{-1} {\cal L}^B = &  \frac{1}{2}  M_P^2  R  
+  \frac{1}{2} M_P^2 {\cal J}''  \p C \p C 
-  \frac{1}{ 2} M_P^4  ({\cal J}')^2 
\\
&- \frac{1}{4}  F^{mn} F_{mn} (v) 
+  \frac{1}{2} M_P^2 {\cal J}'' v^m v_m   
\end{split}
\ee
where
\be
{\cal J}(C) = \frac{3}{2} \text{ln} \ls - \frac{1}{3 M_P^2} {\cal Z}(C) \rs. 
\ee
 For a ghost-free theory one should have 
\be
 {\cal J}'' < 0 . 
\ee
It can be easily seen that the ${\cal J}(C)$ which reproduces the quadratic chaotic model is given by 
\be
\label{chaold}
{\cal J} = - \frac{m^2}{2 M_P^2} C^2  
\ee
and the part of (\ref{OLD}) relevant to inflation will read 
\be
\label{infold}
e^{-1} {\cal L}_{scalar} =  \frac{1}{2}  M_P^2  R   - \frac{1}{2}  \p \psi \p \psi 
-  \frac{m^2}{ 2}   \psi^2 
\ee 
for the inflaton  
\be
\psi = m C  . 
\ee 
Again  this is a single-field inflationary model.

Nevertheless the function (\ref{chaold}) 
does not correspond to a vector superfield self-coupling of the form 
\be
\label{oldmass}
- \frac{1}{2} m^2 \int d^4 \theta E \,  V^2 
\ee
but will come from a more involved function of $V$. 
Indeed, the vector superfield  function inside (\ref{oldminZ}) has to have the form 
\be
\label{expV}
{\cal Z}(V) = - 3 M_P^2  e^{-\frac{m^2}{3 M_P^2} V^2}   
\ee  
which leads to (\ref{chaold}) and (\ref{infold}). 
On the other hand, a quadratic term as (\ref{oldmass}) will give rise to an exponential potential. 
In fact the coupling of the massive vector multiplet to standard supergravity was 
investigated in Ref. \cite{Ferrara:2013rsa}   reproducing, among other models, also quadratic chaotic inflation.

Let us recapitulate for a moment. 
We have shown that starting with a theory of quadratic chaotic inflation we can embed it in supergravity in two distinct ways. 
In one case (the new-minimal) 
there is a softly broken shift symmetry in superspace which can be used to give a technical naturalness argument 
against higher order corrections. 
The embedding can be also straightforwardly carried out in the old-minimal formulation, 
but in this case there is no obvious way to render the theory technically natural.

Finally, one may also interpret (\ref{oldminZ}) as a supergravity 
model with a gauged chiral sector \cite{Wess:1992cp} 
of the form 
\be
\label{oldmingauged}
\begin{split}
{\cal L} = &  \int d^2 \Theta \, 2 {\cal E} \lc  -\frac{1}{8} ( \bar {\cal D}^2 
-8 {\cal R} ) {\cal Z}(\text{ln}[ \bar \Phi e^V \Phi] ) \rc +c.c.
\\
& +  \frac{1}{4} \int d^2 \Theta \, 2 {\cal E} \, W^2(V) + c.c.  
\end{split}
\ee
where the real part of the lowest component of the chiral superfield 
\be
S = \text{ln}\, \Phi
\ee 
will drive inflation, 
while the imaginary will be eaten by the massive vector.

\section{Conclusions}

We have studied the embedding of  the quadratic model of chaotic inflation into the new-minimal and the 
old-minimal theories of supergravity, 
with the use of massive vector multiplets. 
This embedding is quite straightforward and reproduces a single-field inflation model. 
This stems from the underlying Stueckelberg (or BEH) mechanism, 
where the second unwanted component of the scalar multiplet is eaten by the massive vector. 
Indeed this new mechanism has been investigated previously in a series 
of papers \cite{Farakos:2013cqa,Ferrara:2013rsa,Ferrara:2013kca}. 
The simplicity and generality of the models indicate that 
their detailed cosmological properties deserve further study, a problem
to which we will return in the future.

Our additional interest here was the notion of technical naturalness in superspace. 
As we have demonstrated, the model in the new-minimal formulation naturally evades the $\eta$-problem, 
due to a softly broken superspace shift symmetry of the  real linear superfield. 
In the old-minimal formulation, even though the quadratic chaotic model can be successfully embedded, 
we could not identify a corresponding mechanism.

Closing, let  us make a final comment on  the 4D Green-Schwarz terms.  
These were merely introduced to generate a potential for the inflaton, 
and not to cancel some anomaly. 
Nevertheless, 
the specific combination of couplings we have used 
exactly reproduces the 4D analog of the Green-Schwarz mechanism, 
which leads  one to hope that it would be possible to embed the theory into a UV complete superstring model. 
However, in order to protect the form of the potential it is essential that the model exhibits the 
shift symmetry. 
If such a model existed one could extend the technical naturalness of this paper to a top-down naturalness. 
The identification of specific superstring sectors with the supergravity properties discussed in this article 
would offer a new insight to string inflation which we leave for future research.

\section*{Acknowledgements}

We thank I. Dalianis and A. Kehagias  for discussion and correspondence. 
We thank R. Kallosh, A. Linde and M. Porrati for comments on the first version. 
This work is supported by the Grant agency of the Czech republic under the grant P201/12/G028.

\end{document}